\documentclass[twoside,12pt]{article}
\usepackage{amssymb}
\newlength{\dinwidth}
\newlength{\dinmargin}
\newtheorem{theorem}{Theorem}[section]

\def\idty{{\leavevmode\hbox{\rm 1\kern -.3em I}}}

\def\As{{\cal A}}

\def\Hs{{\cal H}}

\def\Js{{\cal J}}

\def\Ls{{\cal L}}
\def\Ms{{\cal M}}
\def\Ns{{\cal N}}
\def\Os{{\cal O}}
\def\Ps{{\cal P}}

\def\Rs{{\cal R}}

\def\Vs{{\cal V}}
\def\Ws{{\cal W}}

\def\idty{{\leavevmode\hbox{\rm 1\kern -.3em I}}}
\def\RR{\textnormal{{\rm I \hskip -5.75pt R}}}
\def\IN{\textnormal{{\rm I \hskip -5.75pt N}}}

\def\CC{\;\textnormal{{\rm \vrule height 6pt width 1pt \hskip -4.5pt C}}}
\def\ZZ{\textnormal{{\rm Z \hskip -8pt Z}}}

\begin{document}
\title{Tomita--Takesaki Modular Theory} 

\author{{\ Stephen J.\ Summers}\\
Department of Mathematics, University of Florida,\\
Gainesville FL 32611, USA\\}

\date{November 15, 2003} 

\maketitle 

\abstract{We provide an brief overview of Tomita--Takesaki modular
theory and some of its applications to mathematical physics. This is
an article commissioned by the {\it Encyclopedia of Mathematical
Physics}, edited by J.-P. Francoise, G. Naber and T.S. Tsun, to be 
published by the Elsevier publishing house.}

\section{Basic Structure}

     The origins of Tomita--Takesaki modular theory lie in two unpublished 
papers of M. Tomita in 1967 and a slim volume \cite{Tak} by M. 
Takesaki. It has developed into one of the most important tools
in the theory of operator algebras and has found many applications in
mathematical physics.

     Though the modular theory has been formulated in a more general
setting, it will be presented in the form in which it most often finds
application in mathematical physics.\footnote{The reader is referred to
\cite{KR,Ped,Str,Tak2} for generalizations, details and references
concerning the material in the first four sections.}  Let $\Ms$ be a
von Neumann algebra on a Hilbert space $\Hs$ containing a
vector $\Omega$ which is cyclic and separating for $\Ms$. Define the
operator $S_0$ on $\Hs$ as follows:
$$S_0 A\Omega = A^* \Omega \quad , \quad \rm{for \; all} 
\quad A \in \Ms \quad .$$
This operator extends to a closed anti-linear operator $S$ defined on
a dense subset of $\Hs$. Let $\Delta$ be the unique positive,
self-adjoint operator and $J$ the unique anti-unitary operator
occurring in the polar decomposition
$$S = J\Delta^{1/2} = \Delta^{-1/2}J \quad . $$
$\Delta$ is called the {\it modular operator} and $J$ the {\it modular
conjugation} (or {\it modular involution}) associated with the pair 
$(\Ms,\Omega)$.  
Note that $J^2$ is the identity operator and $J = J^*$. Moreover, the 
spectral calculus may be applied to $\Delta$ so that $\Delta^{it}$ is a 
unitary operator for each $t \in \RR$ and $\{\Delta^{it} \mid t \in \RR \}$ 
forms a strongly continuous unitary group. Let $\Ms'$ denote the set of
all bounded linear operators on $\Hs$ which commute with all elements
of $\Ms$. The modular theory begins with the following remarkable theorem.

\begin{theorem} \label{TT} 
Let $\Ms$ be a von Neumann algebra with a cyclic and separating 
vector $\Omega$. Then $J\Omega = \Omega = \Delta\Omega$
and the following equalities hold:
$$ J \Ms J = \Ms ' \quad \rm{and} \quad 
\Delta^{it} \Ms \Delta^{-it} = \Ms \quad , \quad \rm{for \; all} 
\quad t \in \RR \quad .$$
\end{theorem}

     Note that if one defines $F_0 A'\Omega = A'^*\Omega$, for all
$A' \in \Ms'$, and takes its closure $F$, then one has the relations
$$ \Delta = FS \quad , \quad \Delta^{-1} = SF \quad , \quad F = J\Delta^{-1/2}
\quad .$$

\section{Modular Automorphism Group}

     By Theorem \ref{TT}, the unitaries $\Delta^{it}, t \in \RR$, induce
a one-parameter automorphism group $\{ \sigma_t \}$ of $\Ms$ by
$$\sigma_t(A) = \Delta^{it} A \Delta^{-it} \quad , \quad A \in \Ms \,
, \, t \in \RR \quad .$$
This group is called the {\it modular automorphism group} of $\Ms$
(relative to $\Omega$). Let $\omega$ denote the faithful normal state on 
$\Ms$ induced by $\Omega$:
$$\omega(A) = \frac{1}{\Vert \Omega \Vert^2} \langle \Omega, A \Omega \rangle
\quad , \quad A \in \Ms \quad .$$
From Theorem \ref{TT} it follows that $\omega$ is invariant under 
$\{ \sigma_t \}$, {\it i.e.} $\omega(\sigma_t(A)) = \omega(A)$ for all 
$A \in \Ms$ and $t \in \RR$.

The modular automorphism group contains information about both $\Ms$ 
and $\omega$. For example, the modular automorphism group is an
inner automorphism on $\Ms$ if and only if $\Ms$ is semi-finite.
It is trivial if and only if $\omega$ is a tracial state on $\Ms$. 
Indeed, one has for any $B \in \Ms$ that $\sigma_t(B) = B$ for all 
$t \in \RR$ if and only if $\omega(AB) = \omega(BA)$ for all $A \in \Ms$.
Let $\Ms^\sigma$ denote the set of all such $B \in \Ms$.

\subsection{The KMS-Condition}

     The modular automorphism group satisfies a condition which had
already been used in mathematical physics to characterize
equilibrium temperature states of quantum systems in statistical
mechanics and field theory --- the {\it Kubo--Martin--Schwinger (KMS)
condition}. If $\Ms$ is a von Neumann algebra and 
$\{\alpha_t \mid t \in\RR\}$ is a $\sigma$-weakly continuous 
one-parameter group of automorphisms of $\Ms$, then the state $\phi$ 
on $\Ms$ satisfies the KMS-condition at (inverse temperature) $\beta$ 
($0 < \beta < \infty$) with respect to $\{\alpha_t\}$ if for any 
$A,B \in \Ms$ there exists a complex function $F_{A,B}(z)$ which is 
analytic on the strip $\{ z \in \CC \mid 0 < {\rm Im}\, z < \beta \}$ 
and continuous on the closure of this strip such that
$$ F_{A,B}(t) = \phi(\alpha_t(A) B) \quad {\rm and} \quad 
F_{A,B}(t + i\beta) = \phi(B \alpha_t(A)) \quad ,$$
for all $t \in \RR$. In this case, 
$\phi(\alpha_{i\beta}(A)B) = \phi(BA)$, for all $A,B$ in a
$\sigma$-weakly dense, $\alpha$-invariant *-subalgebra of $\Ms$. 
Such KMS-states are $\alpha$-invariant, {\it i.e.}
$\phi(\alpha_t(A)) = \phi(A)$, for all $A \in \Ms$, $t \in \RR$, and
are stable and passive (cf. Chapter 5 in \cite{BR2} and \cite{Haag}). 

     Every faithful normal state satisfies the KMS-condition at value
$\beta = 1$ (henceforth called the modular condition) with respect to
the corresponding modular automorphism group.

\begin{theorem} \label{KMS}
Let $\Ms$ be a von Neumann algebra with a cyclic and 
separating vector $\Omega$. Then the induced state $\omega$ on $\Ms$
satisfies the modular condition with respect to the
modular automorphism group $\{ \sigma_t \mid t \in \RR\}$ associated to the
pair $(\Ms,\Omega)$. 
\end{theorem}

     The modular automorphism group is therefore endowed with the
analyticity associated with the KMS-condition, and this is a powerful
tool in many applications of the modular theory to mathematical
physics.  In addition, the physical properties and interpretations of
KMS-states are often invoked when applying modular theory to quantum
physics.

     Note that while the non-triviality of the modular automorphism
group gives a measure of the non-tracial nature of the state, the
KMS-condition for the modular automorphism group provides the missing
link between the values $\omega(AB)$ and $\omega(BA)$, for all $A,B
\in \Ms$ (hence the use of the term ``modular'', as in the theory of
integration on locally compact groups).

     The modular condition is quite restrictive. Only the modular
group can satisfy the modular condition for $(\Ms,\Omega)$, and the
modular group for one state can satisfy the modular condition only in
states differing from the original state by the action of an element
in the center of $\Ms$.

\begin{theorem} Let $\Ms$ be a von Neumann algebra with a cyclic and 
separating vector $\Omega$, and let $\{\sigma_t \}$ be the
corresponding modular automorphism group. If the induced state
$\omega$ satisfies the modular condition with respect to a group
$\{\alpha_t\}$ of automorphisms of $\Ms$, then $\{\alpha_t\}$ must
coincide with $\{\sigma_t \}$. Moreover, a normal state $\psi$ on
$\Ms$ satisfies the modular condition with respect to $\{\sigma_t \}$
if and only if 
$\psi(\cdot) = \omega(h\,\cdot) = \omega(h^{1/2}\,\cdot h^{1/2}) $ 
for some unique positive injective operator $h$ affiliated
with the center of $\Ms$.
\end{theorem}

     Hence, if $\Ms$ is a factor, two distinct states cannot share the
same modular automorphism group. The relation between the
modular automorphism groups for two different states will be described
in more detail.
 
\subsection{One Algebra and Two States}

     Consider a von Neumann algebra $\Ms$ with two cyclic and
separating vectors $\Omega$ and $\Phi$, and denote by $\omega$ and
$\phi$, respectively, the induced states on $\Ms$. Let
$\{\sigma_t^{\omega}\}$ and $\{\sigma_t^{\phi}\}$ denote the
corresponding modular groups.  There is a general relation between the
modular automorphism groups of these states.

\begin{theorem} \label{cocycle}
There exists a $\sigma$-strongly continuous map 
$\RR \ni t \mapsto U_t \in \Ms$ such that

   (1) $U_t$ is unitary, for all $t \in \RR$;

   (2) $U_{t+s} = U_t\sigma_t^{\omega}(U_s)$, for all $s,t \in \RR$;

   (3) $\sigma_t^{\phi}(A) = U_t \sigma_t^{\omega}(A) U_t{}^*$, for all
$A \in \Ms$ and $t \in \RR$.
\end{theorem}

     The one-cocycle $\{ U_t \}$ is commonly called the {\it cocycle
derivative} of $\phi$ with respect to $\omega$ and one writes
$U_t = (D\phi : D\omega)_t$. There is a chain rule for this derivative, as
well: If $\phi, \psi$ and $\rho$ are faithful normal states on $\Ms$,
then $(D\psi : D\phi)_t = (D\psi : D\rho)_t (D\rho : D\phi)_t$, for
all $t \in \RR$. More can be said about the cocycle derivative if the
states satisfy any of the conditions in the following theorem. 

\begin{theorem} \label{equiv} The following conditions are equivalent.

   (1) $\phi$ is $\{\sigma_t^{\omega}\}$-invariant;

   (2) $\omega$ is $\{\sigma_t^{\phi}\}$-invariant;

   (3) there exists a unique positive injective operator $h$ affiliated
with $\Ms^{\sigma^{\omega}} \cap \Ms^{\sigma^{\phi}}$ such that
$\omega( \cdot ) = \phi(h \,\cdot) = \phi(h^{1/2}\,\cdot h^{1/2})$;

   (4) there exists a unique positive injective operator $h'$ affiliated
with $\Ms^{\sigma^{\omega}} \cap \Ms^{\sigma^{\phi}}$ such that
$\phi( \cdot ) = \omega(h' \,\cdot) = \omega(h'^{1/2}\,\cdot h'^{1/2})$;

   (5) the norms of the linear functionals $\omega + i\phi$ and
$\omega - i\phi$ are equal;

   (6) $\sigma_t^{\omega}\sigma_s^{\phi} = \sigma_s^{\phi}\sigma_t^{\omega}$,
for all $s,t \in \RR$.
\end{theorem} 

     The conditions in Theorem \ref{equiv} turn out to be equivalent
to the cocycle derivative being a representation.

\begin{theorem} The cocycle $\{ U_t \}$ intertwining 
$\{\sigma_t^{\omega}\}$ with $\{\sigma_t^{\phi}\}$ is a group
representation of the additive group of reals if and only if $\phi$
and $\omega$ satisfy the conditions in Theorem \ref{equiv}. In that
case, $U(t) = h^{-it}$.
\end{theorem}

     The operator $h' = h^{-1}$ in Theorem \ref{equiv} is called the
{\it Radon--Nikodym derivative} of $\phi$ with respect to $\omega$ and
often denoted by $d\phi/d\omega$, due to the following result, which,
if the algebra $\Ms$ is abelian, {\it is} the well-known
Radon--Nikodym Theorem from measure theory.

\begin{theorem} If $\phi$ and $\omega$ are normal positive linear functionals 
on $\Ms$ such that $\phi(A) \leq \omega(A)$, for all positive elements
$A \in \Ms$, then there exists a unique element $h^{1/2} \in \Ms$ such
that $\phi(\cdot) = \omega(h^{1/2} \cdot h^{1/2})$ and 
$0 \leq h^{1/2} \leq 1$. 
\end{theorem}

     Though there is not sufficient space to treat the matter
properly, the analogies with measure theory are not
accidental. Indeed, any normal trace on a (finite) von Neumann algebra
$\Ms$ gives rise to a noncommutative integration theory in a natural
manner. Modular theory affords an extension of this theory to the
setting of faithful normal functionals $\eta$ on von Neumann algebras
$\Ms$ of any type, enabling the definition of noncommutative $L^p$
spaces, $L^p(\Ms,\eta)$.

\section{Modular Invariants and the Classification of von Neumann Algebras}

     As previously mentioned, the modular structure carries
information about the algebra. This is best evidenced in the structure
of type $III$ factors.  As this theory is rather involved, only a
sketch of some of the results can be given.

     If $\Ms$ is a type $III$ algebra, then its crossed product
$\Ns = \Ms \rtimes_{\sigma^\omega} \RR$ relative to the modular
automorphism group of any faithful normal state $\omega$ on $\Ms$
is a type $II_\infty$ algebra with a faithful semifinite normal
trace $\tau$ such that $\tau \circ \theta_t = e^{-t}\tau$, 
$t \in \RR$, where $\theta$ is the dual of $\sigma^\omega$ on
$\Ns$. Moreover, the algebra $\Ms$ is isomorphic to the
cross product $\Ns \rtimes_\theta \RR$, and this decomposition is
unique in a very strong sense. This structure theorem entails the
existence of important algebraic invariants for $\Ms$, which has many
consequences, one of which is made explicit here.

     If $\omega$ is a faithful normal state of a von Neumann algebra $\Ms$
induced by $\Omega$, let $\Delta_{\omega}$ denote the modular operator 
associated to $(\Ms,\Omega)$ and ${\rm sp} \,\Delta_{\omega}$ denote the 
spectrum of $\Delta_{\omega}$. The intersection 
$$S'(\Ms) = \cap \; {\rm sp} \,\Delta_{\omega} $$
over all faithful normal states $\omega$ of $\Ms$ is an algebraic
invariant of $\Ms$. 

\begin{theorem} \label{spectrum}
Let $\Ms$ be a factor acting on a separable Hilbert space. If $\Ms$ is
of type $III$, then $0 \in S'(\Ms)$; otherwise, $S'(\Ms) = \{ 0,1 \}$
if $\Ms$ is of type $I_\infty$ or $II_\infty$ and $S'(\Ms) = \{ 1 \}$
if not. Let $\Ms$ now be a factor of type $III$.

     (i) $\Ms$ is of type $III_\lambda$, $0 < \lambda < 1$, if and only
if $S'(\Ms) = \{ 0 \} \cup \{ \lambda^n \mid n \in \ZZ \}$.

     (ii) $\Ms$ is of type $III_0$ if and only if $S'(\Ms) = \{ 0,1 \}$.

     (iii) $\Ms$ is of type $III_1$ if and only if $S'(\Ms) = [0,\infty)$.

\end{theorem}

     In certain physically relevant situations, the spectra of the
modular operators of all faithful normal states coincide, so that 
Theorem \ref{spectrum} entails that it suffices to compute the 
spectrum of any conveniently chosen modular operator in order to 
determine the type of $\Ms$. In other such situations, there are 
distinguished states $\omega$ such that $S'(\Ms) = {\rm sp} \,\Delta_\omega$. 
One such example is provided by asymptotically abelian systems. A von Neumann
algebra $\Ms$ is said to be {\it asymptotically abelian} if there
exists a sequence $\{\alpha_n\}_{n \in \IN}$ of automorphisms of $\Ms$
such that the limit of $\{ A \alpha_n(B) - \alpha_n(B)A\}_{n\in\IN}$ 
in the strong operator topology is zero, for all $A,B \in \Ms$. If the state
$\omega$ is $\alpha_n$-invariant, for all $n \in \IN$, then 
${\rm sp} \,\Delta_{\omega}$ is contained in ${\rm sp} \,\Delta_\phi$, 
for all faithful normal states $\phi$ on $\Ms$, so that 
$S'(\Ms) = {\rm sp} \,\Delta_\omega$. 
If, moreover, ${\rm sp} \,\Delta_\omega = [0,\infty)$, then 
${\rm sp} \,\Delta_\omega = {\rm sp} \,\Delta_\phi$, for all $\phi$ 
as described. 

\section{Self-Dual Cones}

     Let $j : \Ms \rightarrow \Ms'$ denote the antilinear *-isomorphism
defined by $j(A) = JAJ$, $A \in \Ms$. The {\it natural positive cone}
$\Ps^\natural$ associated with the pair $(\Ms,\Omega)$ is defined as the
closure in $\Hs$ of the set of vectors
$$\{ Aj(A)\Omega  \mid A \in \Ms \} \quad . $$

\noindent  Let $\Ms_+$ denote the set of all positive elements of $\Ms$.
The following theorem collects the main attributes of the natural cone.

\begin{theorem} 
(1) $\Ps^\natural$ coincides with the closure in $\Hs$ of the set 
$\{ \Delta^{1/4}A\Omega \mid A \in \Ms_+ \}$.

(2) $\Delta^{it} \Ps^\natural = \Ps^\natural$ for all $t \in \RR$.

(3) $J\Phi = \Phi$ for all $\Phi \in \Ps^\natural$.

(4) $Aj(A)\Ps^\natural \subset \Ps^\natural$ for all $A \in \Ms$.

(5) $\Ps^\natural$ is a pointed, self-dual cone whose linear span 
coincides with $\Hs$.

(6) If $\Phi \in \Ps^\natural$, then $\Phi$ is cyclic for $\Ms$ if and only if
$\Phi$ is separating for $\Ms$.

(7) If $\Phi \in \Ps^\natural$ is cyclic, and hence separating, for $\Ms$,
then the modular conjugation and the natural cone associated 
with the pair $(\Ms,\Phi)$ coincide with $J$ and $\Ps^\natural$, respectively.

(8) For every normal positive linear functional $\phi$ on $\Ms$ there
exists a unique vector $\Phi_{\phi} \in \Ps^\natural$ such that 
$\phi(A) = \langle\Phi_{\phi}, A \Phi_{\phi}\rangle$, for all $A \in \Ms$. 
\end{theorem}

     In fact, the algebras $\Ms$ and $\Ms'$ are uniquely characterized
by the natural cone $\Ps^\natural$ \cite{Connes}. In light of (8),
if $\alpha$ is an automorphism of $\Ms$, then
$$V(\alpha)\Phi_{\phi} = \Phi_{\phi \circ \alpha^{-1}}$$
defines an isometric operator on $\Ps^\natural$, which by (5) extends 
to a unitary operator on $\Hs$. The map $\alpha \mapsto V(\alpha)$ 
defines a unitary representation of the group of automorphisms 
${\rm Aut}(\Ms)$ on $\Ms$ in such a manner that 
$V(\alpha)AV(\alpha)^{-1} = \alpha(A)$ for all 
$A \in \Ms$ and $\alpha \in {\rm Aut}(\Ms)$. Indeed, one has the following.

\begin{theorem}
Let $\Ms$ be a von Neumann algebra with a cyclic and separating vector 
$\Omega$. The group $\Vs$ of all unitaries $V$ satisfying 
$$ V \Ms V^* = \Ms \quad , \quad VJV^* = J \quad , \quad 
V\Ps^\natural = \Ps^\natural$$
is isomorphic to ${\rm Aut}(\Ms)$ under the above map 
$\alpha \mapsto V(\alpha)$, which is called the {\rm standard implementation} 
of ${\rm Aut}(\Ms)$.
\end{theorem}

     Often of particular physical interest are (anti-)automorphisms of
$\Ms$ leaving $\omega$ invariant. They can only be implemented by
(anti-)unitaries which leave the pair $(\Ms,\Omega)$ invariant. In
fact, if $U$ is a unitary or anti-unitary operator satisfying $U
\Omega = \Omega$ and $U \Ms U^* = \Ms$, then $U$ commutes with both
$J$ and $\Delta$.

\section{Two Algebras and One State} \label{inclusions}

     Motivated by applications to quantum field theory, the study of 
the modular structures associated with one state and more than one von Neumann
algebra has begun (see the review paper \cite{Bor} for references and
details).  Let $\Ns \subset \Ms$ be von Neumann algebras with a
common cyclic and separating vector $\Omega$. $\Delta_{\Ns}, J_{\Ns}$
and $\Delta_{\Ms}, J_{\Ms}$ will denote the corresponding modular
objects. The structure $(\Ms,\Ns,\Omega)$ is called a
{\it$\pm$-half-sided modular inclusion} if $\Delta_{\Ms}^{it} \Ns
\Delta_{\Ms}^{-it} \subset \Ns$, for all $\pm t \geq 0$. 

\begin{theorem} \label{Wiesbrock}
Let $\Ms$ be a von Neumann algebra with cyclic and separating vector
$\Omega$. The following are equivalent.

   (i) There exists a proper subalgebra $\Ns \subset \Ms$ such that 
$(\Ms,\Ns,\Omega)$ is a $\mp$-half-sided modular inclusion.

   (ii) There exists a unitary group $\{ U(t) \}$ with positive generator
such that 
$$U(t) \Ms U(t)^{-1} \subset \Ms \, , \, {\rm for \, all} \; \pm t \geq 0 ,
\quad {\rm and} \quad U(t)\Omega = \Omega \, , \, {\rm for \, all} \; 
t \in \RR \quad .$$
Moreover, if these conditions are satisfied, then the following relations must
hold:
$$\Delta_{\Ms}^{it} U(s) \Delta_{\Ms}^{-it} = 
\Delta_{\Ns}^{it} U(s) \Delta_{\Ns}^{-it} = U(e^{\mp 2\pi t}s)$$
and
$$J_{\Ms} U(s) J_{\Ms} = J_{\Ns} U(s) J_{\Ns} = U(-s) \quad , $$
for all $s,t \in \RR$. In addition, $\Ns = U(\pm 1) \Ms U(\pm 1)^{-1}$,
and if $\Ms$ is a factor, it must be type $III_1$.
\end{theorem} 

     The richness of this structure is further suggested by the 
next theorem.

\begin{theorem} \label{Wiesbrock2}
   (a) Let $(\Ms,\Ns_1,\Omega)$ and $(\Ms,\Ns_2,\Omega)$ be
$-$-half-sided, resp. $+$-half-sided, modular inclusions satisfying
the condition $J_{\Ns_1}J_{\Ns_2} = J_{\Ms}J_{\Ns_2}J_{\Ns_1}J_{\Ms}$.
Then the modular unitaries 
$\Delta_{\Ms}^{it},\Delta_{\Ns_1}^{is},\Delta_{\Ns_2}^{iu}$,
$s,t,u \in \RR$, generate a faithful continuous unitary representation
of the identity component of the group of isometries of two-dimensional
Minkowski space.

   (b) Let $\Ms,\Ns, \Ns\cap\Ms$ be von Neumann algebras with a common
cyclic and separating vector $\Omega$. If $(\Ms,\Ms\cap\Ns,\Omega)$
and $(\Ns,\Ms\cap\Ns,\Omega)$ are $-$-half-sided, resp. $+$-half-sided,
modular inclusions such that $J_\Ns \Ms J_\Ns = \Ms$, then
the modular unitaries 
$\Delta_{\Ms}^{it},\Delta_{\Ns}^{is},\Delta_{\Ns \cap \Ms}^{iu}$,
$s,t,u \in \RR$, generate a faithful continuous unitary representation
of $SL(2,\RR)/\ZZ_2$.
\end{theorem} 

     This has led to a further useful notion. If $\Ns \subset \Ms$ and
$\Omega$ is cyclic for $\Ns \cap \Ms$, then $(\Ms,\Ns,\Omega)$ is said
to be a {\it $\pm$-modular intersection} if both
$(\Ms,\Ms\cap\Ns,\Omega)$ and $(\Ns,\Ms\cap\Ns,\Omega)$ are
$\pm$-half-sided modular inclusions and
$$J_\Ns \big[ \lim_{t \rightarrow \mp \infty} 
  \Delta_{\Ns}^{it}\Delta_{\Ms}^{-it} \big] J_\Ns = 
  \lim_{t \rightarrow \mp \infty} \Delta_{\Ms}^{it}\Delta_{\Ns}^{-it} \, ,$$
where the existence of the strong operator limits is assured by the
preceding assumptions. An example of the utility of this structure 
is the following theorem.

\begin{theorem} \label{Wiesbrock3}
Let $\Ns,\Ms,\Ls$ be von Neumann algebras with a common cyclic and
separating vector $\Omega$. If $(\Ms,\Ns,\Omega)$ and $(\Ns',\Ls,\Omega)$
are $-$-modular intersections and $(\Ms,\Ls,\Omega)$ is a $+$-modular
intersection, then the unitaries 
$\Delta_{\Ms}^{it},\Delta_{\Ns}^{is},\Delta_{\Ls}^{iu}$,
$s,t,u \in \RR$, generate a faithful continuous unitary representation
of $SO^\uparrow(1,2)$.
\end{theorem}

     These results and their extensions to larger numbers of algebras
were developed for application in algebraic quantum field theory, but
one may anticipate that half-sided modular inclusions will find wider
use. Modular theory has also been fruitfully applied in the theory of
inclusions $\Ns \subset \Ms$ of properly infinite algebras with finite
or infinite index.

\section{Applications in Quantum Theory}

     Tomita--Takesaki theory has found many applications in quantum
field theory and quantum statistical mechanics. As previously
mentioned, the modular automorphism group satisfies the KMS-condition,
a property of physical significance in the quantum theory of
many-particle systems, which includes quantum statistical mechanics
and quantum field theory. In such settings it occurs that for a
suitable algebra of observables $\Ms$ and state $\omega$ an
automorphism group $\{ \sigma_{\beta t}\}$ representing the time
evolution of the system satisfies the modular condition. Hence, on the
one hand, $\{ \sigma_{\beta t}\}$ is the modular automorphism group of
the pair $(\Ms,\Omega)$, and, on the other hand, $\omega$ is an
equilibrium state at inverse temperature $\beta$, with all the
consequences which both of these facts have.

     But it has become increasingly clear that the modular objects
$\Delta^{it}$, $J$, of certain algebras of observables and states
encode additional physical information.  In 1975, it was discovered
that if one considers the algebras of observables associated with a
finite-component quantum field theory satisfying the Wightman axioms,
then the modular objects associated with the vacuum state and algebras
of observables localized in certain wedge-shaped regions in Minkowski
space have geometric content. In fact, the unitary group $\{
\Delta^{it} \}$ implements the group of Lorentz boosts leaving the
wedge region invariant (this property is now called {\it modular
covariance}), and the modular involution $J$ implements the spacetime
reflection about the edge of the wedge, along with a charge
conjugation. This discovery caused some intense research
activity.\footnote{See \cite{BW,Bor,Haag} for further details and
references.}

\subsection{Positive Energy}

     In quantum physics the time development of the system is often
represented by a strongly continuous group $\{ U(t) = e^{itH} \mid t
\in \RR\}$ of unitary operators, and the generator $H$ is interpreted
as the total energy of the system. There is a link between modular
structure and positive energy, which has found many applications in
quantum field theory. This result was crucial in the development of
Theorem \ref{Wiesbrock} and was motivated by the 1975 discovery
mentioned above, now commonly called the Bisognano--Wichmann Theorem.
 
\begin{theorem} \label{positive}
Let $\Ms$ be a von Neumann algebra with a cyclic and 
separating vector $\Omega$, and let $\{ U(t) \}$ be a continuous
unitary group satisfying $U(t) \Ms U(-t) \subset \Ms$, for all 
$t \geq 0$. Then any two of the following conditions imply the third.

   (1) $U(t) = e^{itH}$, with $H \geq 0$;

   (2) $U(t) \Omega = \Omega$, for all $t \in \RR$;

   (3) $\Delta^{it} U(s) \Delta^{-it} = U(e^{-2\pi t}s)$ and
$JU(s)J = U(-s)$, for all $s,t \in \RR$.

\end{theorem}

\subsection{Modular Nuclearity and Phase Space Properties}

     Modular theory can be used to express physically meaningful
properties of quantum ``phase spaces'' by a condition of compactness
or nuclearity of certain maps.  In its initial form, the condition was
formulated in terms of the Hamiltonian, the global energy operator of
theories in Minkowski space. The above indications that the modular
operators carry information about the energy of the system were
reinforced when it was shown that a formulation in terms of modular
operators was essentially equivalent.

     Let $\Os_1 \subset \Os_2$ be nonempty bounded open subregions of
Minkowski space with corresponding algebras of observables
$\As(\Os_1)\subset\As(\Os_2)$ in a vacuum representation with vacuum
vector $\Omega$, and let $\Delta$ be the modular operator associated
with $(\As(\Os_2),\Omega)$ (by the Reeh--Schlieder Theorem, $\Omega$
is cyclic and separating for $\As(\Os_2)$). For each 
$\lambda \in (0,1/2)$ define the mapping 
$\Xi_{\lambda} : \As(\Os_1) \rightarrow \Hs$ by 
$\Xi_{\lambda}(A) = \Delta^{\lambda}A\Omega$. The compactness of any one
of these mappings implies the compactness of all of the others. Moreover,
the $l^p$ (nuclear) norms of these mappings are interrelated and provide
a measure of the number of local degrees of freedom of the system.
Suitable conditions on the maps in terms of these norms
entail the strong statistical independence condition called the 
split property. Conversely, the split property implies the compactness
of all of these maps. Moreover, the existence of equilibrium temperature 
states on the global algebra of observables can be derived from suitable 
conditions on these norms in the vacuum sector. 

     The conceptual advantage of the modular compactness and nuclearity
conditions compared to their original Hamiltonian form lies in the fact
that they are meaningful also for quantum systems in curved space--times, 
where global energy operators ({\it i.e.} generators corresponding to
global timelike Killing vector fields) need not exist.

\subsection{Modular Position and Quantum Field Theory}

     The characterization of the relative ``geometric'' position of 
algebras based on the notions of modular inclusion and modular 
intersection was directly motivated by the Bisognano-Wichmann Theorem. 
Observable algebras associated with suitably chosen wedge regions in
Minkowski space provided examples whose essential structure could be
abstracted for more general application, resulting in the notions presented 
in Section \ref{inclusions}.  

     Theorem \ref{Wiesbrock2}(b) has been used to construct from two
algebras and the indicated half-sided modular inclusions a conformal
quantum field theory on the circle (compactified light ray) with
positive energy. And since the chiral part of a conformal quantum
field model in two spacetime dimensions naturally yields such
half-sided modular inclusions, studying the inclusions in Theorem
\ref{Wiesbrock2}(b) is equivalent to studying such field
theories. Theorems \ref{Wiesbrock2}(a) and \ref{Wiesbrock3} and their
generalizations to inclusions involving up to 6 algebras have been
employed to construct Poincar\'e covariant nets of observable algebras
(the algebraic form of quantum field theories) satisfying the spectrum
condition on $d+1$-dimensional Minkowski space for $d = 1,2,3$.
Conversely, such quantum field theories naturally yield such systems
of algebras.

     This intimate relation would seem to open up the possibility of
constructing interacting quantum field theories from a limited number
of modular inclusions/intersections.

\subsection{Geometric Modular Action}

     The fact that the modular objects in quantum field theory
associated with wedge-shaped regions and the vacuum state 
in Minkowski space have geometric
significance (``geometric modular action'') was originally discovered
in the framework of the Wightman axioms. As algebraic quantum field
theory (AQFT) does not rely on the concept of Wightman fields, it was
natural to ask (1) when does geometric modular action hold in AQFT and
(2) which physically relevant consequences follow from this feature?

     There are two approaches to the study of geometric modular
action.  In the first, attention is focused on modular covariance,
expressed in terms of the modular groups associated with wedge
algebras and the vacuum state in Minkowski space. Modular covariance
has been proven to obtain in conformally invariant AQFT, in any
massive theory satisfying asymptotic completeness, and also in the
presence of other, physically natural assumptions. To mention only
three of its consequences, both the Spin--Statistics Theorem and the
PCT Theorem, as well as the existence of a continuous unitary
representation of the Poincar\'e group acting covariantly upon the
observable algebras and satisfying the spectrum condition 
follow from modular covariance.

     In a second approach to geometric modular action
the modular involutions are the primary focus. Here, no {\it a priori} 
connection between the modular objects and isometries of the space--time 
is assumed. The central assumption, given the state vector $\Omega$
and the von Neumann algebras of localized observables $\{ \As(\Os) \}$ 
on the space--time, is that there exists a family $\Ws$ of subsets of 
the space--time such that 
$J_{W_1} \Rs(W_2) J_{W_1} \in \{ \Rs(W) \mid W\in\Ws\}$, for every
$W_1,W_2 \in \Ws$.
This condition makes no explicit appeal to isometries or other special
attributes and is thus applicable in principle to quantum field theories 
on general curved space--times.  

     It has been shown for certain space--times,
including Minkowski space, that under certain additional technical 
assumptions, the modular involutions encode enough information to 
determine the dynamics of the theory, the isometry group of the space--time, 
and a continuous unitary representation of the isometry group which acts 
covariantly upon the observables and leaves the state invariant. 
In certain cases including Minkowski space, it is even possible to derive 
the space--time itself from the group $\Js$ generated by the modular 
involutions $\{ J_W \mid W \in \Ws \}$. 

     The modular unitaries $\Delta_W^{it}$ enter in this approach
through a condition which is designed to assure the stability of the
theory, namely that $\Delta_W^{it} \in \Js$, for all $t \in \RR$ and
$W \in \Ws$.  In Minkowski space this additional condition entails
that the derived representation of the Poincar\'e group satisfies the
spectrum condition.

\subsection{Further Applications}

     As previously observed, through the close connection to
the KMS condition, modular theory enters naturally into the equilibrium
thermodynamics of many-body systems. But in recent work on the theory of
nonequilibrium thermodynamics it also plays a role in making
mathematical sense of the notion of quantum systems in local thermodynamic
equilibrium. Modular theory has also proved to be of utility in 
recent developments in the theory of superselection rules and their
attendant sectors, charges and charge-carrying fields.

\end{document}